\documentstyle[12pt]{article}
\begin{document}
%\pagenumbering{arabic}
\title{
%===== please type here your talk title using capital letters ======
QUANTUM FIELD THEORY AND THE HELIUM ATOM: 101 YEARS LATER
\footnote{Invited talk: ``Quantum Systems:
New Trends and Methods", Minsk, June 3-7, 1996}\\}
%===================================================================
%{\small (For 20\% Reduction to 8.5in$\times$6in Trim Size)}}
\author{
%=============== first author's name & his address =============
      J. SUCHER \\
{\it Department of Physics, University of Maryland} \\
{\it College Park, MD 20742, U.S.A.}}
%%==============================================================
%\and
%=============== second author's name & his address ============
%      SECOND AUTHOR'S NAME \\
%{\it Group, Company, Address, City, State ZIP/Zone, Country}}
%%==============================================================
\maketitle

\begin{abstract}
Helium was first isolated on Earth in 1895, by Sir William Ramsey.
One hundred and one years later, it seems like a good time to review
our current theoretical understanding of the helium atom.  Helium has
played an important role both in the development of quantum mechanics
and in quantum field theory.  The early history of helium is sketched.
Various aspects of the modern theory are described in some detail,
including (1)~the computation of fine structure to order $\alpha^2 Ry$ and
$\alpha^3 Ry$, (2) the decay of metastable states, and (3) Rydberg states and
long-range forces.  A brief survey is made of
some of the recent work on the quantum field theory of $He$ and
$He$-like ions.

\end{abstract}

\newpage
\section{Introduction}
The title of my talk is inspired by that of a famous novel by
Gabriel Garcia Marqu$\acute{e}$z: ``100 years of solitude."  Helium,
whose existence was not even suspected till the middle of the 19th century,
has experienced exactly 101 years of attention since its terrestrial
discovery in 1895.  It has played a major role in the development
of atomic, nuclear, and condensed matter physics.  The helium atom,
and its cousins, the $He$-like ions, continue to be a subject of
active study.  It has played an important role, second only to hydrogen,
both in the development of quantum mechanics and that of quantum field
theory (QFT).  Since this atom is, in the sense of the title of this
conference, a quantum system {\it par excellence},
I thought this would be an appropriate time to review the
history of this subject, with emphasis on the QFT aspects.

I will begin (Sec.~2) with a historical sketch concerning the discovery of
helium, first in the Sun and then on Earth, including some fascinating
facts which I gleaned by browsing through a wonderful book by Isaac
Asimov.\cite{asim}  (I take special pleasure in referring to Asimov,
for several reasons:  He was born in Belarus (Petrovici, 1920)
and he is a distant relative of one of the members of the organizing committee!
Moreover, he obtained his higher degrees at Columbia University, where
I was a student many years later.)  I will then describe the importance
of the role played by study of $He$ in the development of quantum
mechanics (Sec.~3) and in the early stages of quantum electrodynamics
(QED) (Sec.~4).  I then turn
to developments beginning in the 1950's, including the first calculation
of atomic energy levels to accuracy $\alpha^3 Ry$ (Sec.~5).  Next, I
will consider the theory of the decays of excited states of $He$ and
$He$-like ions, especially of metastable states, in which the Sun enters
our story once again (Sec.~6).   In Sec.~7, I note the role
played by study of high-lying, so-called Rydberg states (n $\geq$ 10) of
$He$ in advancing the theory of long-range forces arising from two-photon
exchange. I conclude in Sec.~8 with a brief survey of some of the recent work
involving QED and $He$-like systems.

\section{The discovery of helium:  From Janssen to Ramsay}
The story begins with someone whom you have probably never heard of,
one P.J.C. Janssen (1822-1907).  From his name you might
think he is yet another
one of those Scandinavians who have contributed so much to the discovery
of the elements ... but you would be wrong.  The initials stand for
Pierre Jules Cesar and his name is pronounced in the French manner
(Zhahn-sen').
Born in Paris, Janssen was a great traveler in the interests
of astronomy.  He went to Peru and the Azores, to Italy and Greece, to Japan
and Siam.  In 1868, still without benefit of United/Lufthansa and the like,
he traveled to India to observe a total eclipse of the sun.
More about that in a moment.
But to give you an idea of Janssen's daring and determination,
I add that in 1870, when Paris was besieged by the Prussians,
and Janssen was eager to get to Algeria to observe another total eclipse,
he escaped from Paris by balloon flight!
During his observations in India, Janssen noticed a strange spectral line
and forwarded his data to Sir Norman Joseph Lockyer (1836-1920), a
British astronomer who was a pioneer in the study of solar spectra and
coined the term ``chromosphere".  Lockyer compared the
position of Janssen's line to lines of known elements and
concluded that it must belong to an as yet unknown element,
possibly not even existing on earth.  He named it helium,
after the Greek work for the sun, ``helios".  But his suggestion was
dismissed by chemists, who were not prepared to accept the
existence of a new element on what was then regarded
(not without reason) as a flimsy basis.  Indeed subsequently,
many new lines were discovered and incorrectly attributed to new elements,
such as ``coronium", ``nebulium," and so on;
all of these turned out to correspond to ionized states of known elements.

Our story now turns to the two R's, Rayleigh and Ramsay.
Lord Rayleigh (1842-1919) is known to most of us for his work in
mathematical physics.  But he was also a great experimentalist.
During an extensive series of measurements on the density of gases,
undertaken as a test of Prout's hypothesis, he noted that the density
of nitrogen obtained from the air appeared to be very slightly
higher than that of the nitrogen obtained from chemical
compounds, unlike the case with oxygen.
In frustration he wrote to the journal {\it Nature},
(founded in 1869 by Lockyer himself and still going strong) for suggestions.
In response, after getting permission from Rayleigh to do so (!),
the matter was taken up by Sir William Ramsay (1852-1916) in 1892.
Ramsay remembered that a century earlier
Cavendish had found that when he tried to combine the nitrogen of
the air with oxygen a final bubble of air was left over, which Cavendish
speculated to be a gas heavier than nitrogen which did not combine with oxygen.
Ramsay repeated the experiment of Cavendish and found a similar bubble.
Using the now available spectroscope, Ramsay and Rayleigh studied the
lines from the left-over gas and identified for the first time a
zero-valence element, which they called ``argon," from the Greek word for inert.
With the periodic table as a guide, Ramsay guessed that there might be a
whole column of inert elements and began the search.  In 1895, following up
earlier work of the American geologist W. Hillebrand, he isolated a gas from a
uranium mineral called clevite, previously mistaken for nitrogen, which
showed the same bright spectral line seen by Janssen over 25 years before.
Helium had been found on Earth!  Later, starting with liquid argon,
Ramsay isolated three more inert gases:
neon (new), krypton (hidden) and xenon (stranger).  In 1904, Ramsay and
Rayleigh received the Nobel Prizes for chemistry and
physics, respectively, for their work on the inert gases.

\section{Helium and quantum mechanics:  From Bohr to Hylleraas}
Within a decade or so after its discovery, helium began to be of great
importance in the development of atomic, nuclear and condensed matter physics.
  With regard to condensed-matter physics,
in 1908 Kammerlingh-Onnes liquified helium at 4 K; in 1911 he used helium
in the discovery of superconductivity.  With regard to nuclear physics,
helium was in at the beginning:  Rutherford's discovery of the first
nucleus, that of gold, involved an $\alpha$-particle beam and in 1906-1909
Geiger and Rutherford identified the $\alpha$-particle as the doubly ionized
helium atom.

The discovery of the atomic nucleus led directly to the Bohr
model and to Bohr's explanation of the spectrum of hydrogen in
terms of quantization of the action.  However, the attempt
by Bohr and Sommerfeld to extend the model to account for the
spectrum of helium failed.  In the 1920's this failure led in
part to the discovery of quantum mechanics by Heisenberg and
Schr\"{o}dinger and, in particular, to the Schr\"{o}dinger equation
for the stationary states of the hydrogen atom or hydrogen-like ions:
\begin{equation}
h(1)\phi (1) = [\frac{{\bf p}_{1}^{2}}{2m} + U_{C}(1)] \phi(1) = W\phi (1).
\label{hamil}
\end{equation}
with $U_{C}(1)$ the Coulomb interaction of the electron with a point
nucleus: $U_{C}(1) = -Ze^{2}/4\pi r_{1}.$
With the constraint that $\phi$ is normalizable,
one recovers the Bohr levels as eigenvalues
of $h(1)$.

What about helium?  Within the Schr\"{o}dinger framework the energy levels
ought to be determined by
\begin{equation}
h(1,2)\phi (1,2) = W\phi (1,2),
\label{3.2}
\end{equation}
where
\begin{equation}
h(1,2) = h(1) + h(2) + U_{C} (1,2),
\label{3.3}
\end{equation}
with $U_{C} (1,2) = e^{2}/4\pi r_{12}$.
Since $(\ref{3.2})$ can no longer be solved analytically, approximation
methods must be used.  This problem was attacked in the later 1920's
by several authors but most successfully by Egil Hylleraas
\cite{hyll} who,
many years before Mandelstam, introduced a set of {\it s},
{\it t} and {\it u} variables
into physics which were to become famous:
\begin{equation}
        s = r_{1}+r_{2}, \; t = r_{1}-r_{2}, \; u = r_{12} .
\label{3.4}
\end{equation}
For spherically symmetric $(L = 0)$ states, the spatial wave function
$\phi (1,2)$ may be taken to depend only on these rotationally invariant
variables.  As an ansatz for the application of the variational principle
\begin{equation}
        \delta <\phi \mid h(1,2) \mid \phi> = 0,
\label{3.5}
\end{equation}
Hylleraas wrote
\begin{equation}
        \phi = \phi (s,t,u) = f(ks,kt,ku).
\label{3.6}
\end{equation}
Here $k$ is a "nonlinear parameter" and $f$ is of the form [in atomic
units]
\begin{equation}
        f(s,t,u) = e^{-s/2} P(s,t,u),
\label{3.7}
\end{equation}
with P(s,t,u) a multinomial in s, t, and u:
\begin{equation}
        P(s,t,u) = \Sigma c_{n,2l,m} s^n t^{2l} u^m
\label{3.8}
\end{equation}
involving "linear parameters" $c_{ijk}$.  With a six-parameter {\it P},
his result for the energy of the helium ground state gave
for the ionization energy, $J_{th} = 1.8065$ Ry, to be compared with
the experimental value at that time, $J_{exp} = 1.807$ Ry.
This agreement, to better than a part in $10^3$, was important in
putting the approach to atomic spectra based on a many-body
Schr\"{o}dinger equation on firm, quantitative ground.

In the subsequent 60 years, and especially after the advent
of modern computers, the increasingly precise study of the discrete spectrum
of the non-relativistic Hamiltonian (\ref{hamil})
developed into a small industry.  As I will describe later, in recent
years this study has in fact taken a great leap forward.  The obvious
generalization of $(\ref{3.3})$ to an N-electron atom or ion $(N > 2)$ ,
when supplemented by the antisymmetry principle and electron spin wave
functions, constitutes the foundation of the nonrelativistic theory of
atomic spectra.

\section{Helium and QED:  From Darwin to Breit}

\subsection{\it The patchwork era}

Of course, no
matter how accurately $(\ref{3.3})$ is solved, the energy
levels obtained from it are not accurate enough for comparison with very
precise experiments.  Even in classical electrodynamics (CED), the
interaction between two point-like charged particles is not given just
by their Coulomb interaction, which is all that is included in
$(\ref{3.3})$.  Already in 1920, Darwin\cite{darwin} had shown how
the inclusion of
relativistic effects in CED leads, to order $v^2 /c^2$ , to an effective
Lagrangian interaction term of the form
\begin{equation}
L_{int} = \frac{(e_1 e_2/4 \pi c^2) ({\bf v}_1 \cdot {\bf v}_2 +
{\bf v}_1 \cdot {\hat {\bf r}} {\bf v}_2 \cdot
{\bf \hat{r}})}{2r}.
\label{4.1}
\end{equation}
Upon quantization, this leads to a version of what later came to be
called the {\bf orbit-orbit interaction}, $U_{o-o}$.
Further, after the discovery of electron spin and the
associated magnetic moment, it was clear that to describe the
fine-structure accurately one would have to include not only the
{\bf spin-self-orbit interaction} $U_{s-s-o}$ (i) of each electron with the
nucleus, as in the case of hydrogen, but also a {\bf spin-other-orbit
interaction} $U_{s-o-o}$ (1,2) and a {\bf spin-spin interaction} $U_{s-s}
(1,2)$.  These terms could be, and were, written down by analogy with the
corresponding classical terms, although not completely correctly in the
case of $U_{s-s}$.  It was also clear that at this level of accuracy
$(\alpha^2 Ry)$ one should include, again as in H, the relativistic
kinetic energy correction $U_{kin}$ proportional to ${\bf p}^4_i/m^3$, for
each electron "i".

Is that all?  The discovery of the (one-electron) Dirac equation led one
to expect additional terms.  Reduction to nonrelativistic form of the
Dirac equation for H or an H-like ion, viz.
\begin{equation}
(h_D(1) \, + \, U_C(1)){\bf \psi} \,  = \, E{\bf \psi}, \quad [h_D(1) \equiv
{\bf \alpha}_1 \cdot {\bf p}_1 + \beta_1 m]
\label{4.2}
\end{equation}
leads not only to the spin-orbit and kinetic energy corrections but also
to a contact term $U_{cont}(1)$, the so-called "S-state interaction",
which has no classical analogue.  Clearly, there must be counterparts of
this in the case of helium.  In summary, as a result of all this
patchwork, one expects the leading correction to the energy levels
obtained from $(\ref{3.2})$ to take the form
\begin{equation}
W^{(2)} \, = \, <{\bf \psi}_{SP} \mid U_{eff}^{(2)} \mid {\bf \psi}_{SP}>
\label{4.3}
\end{equation}
where ${\bf \psi}_{SP}(1,2)$ is a Schr\"{o}dinger-Pauli (SP) type wave
function and
\begin{equation}
U_{eff}^{(2)} \, = \, U_{s.i.} \, + \, U_{s.d.}
\label{4.4}
\end{equation}
with $U_{s.i.}$ and $U_{s.d.}$ the spin-independent and spin-dependent
parts of the effective interaction $U_{eff}^{(2)}$:
\begin{equation}
U_{s.i.} \, = \, \Sigma [U_{kin} (i) + U_{cont} (i)] \, + \, U_{o-o}
(1,2) ,
\end{equation}
\begin{equation}
U_{s.d.} \, = \, \Sigma U_{s-o} (i) \, + \, U_{s-o-o}(1,2) \, + \,
U_{s-s} (1,2) .
\end{equation}
I will write down the explicit expressions later [Sec. 5.2].

\subsection{\it The next stage}

Even if the interaction between two electrons were perfectly described
by $U_C (r_{12})$ , it was evident that to take into account relativistic
and spin-dependent effects in a unified and systematic way it would be
desirable to construct a generalization of the one-electron Dirac
equation to two electrons.  The first steps in this direction and in
improving the above patchwork approach were taken by J.B.
Gaunt\cite{gaunt} and especially by Gregory Breit.\cite{breit}

\subsubsection{The two-electron Dirac-Coulomb Hamiltonian.}
As a starting point for a relativistic description of the $He$ atom, Breit
took what appears to be an obvious extension of (\ref{4.2}):  He
replaced the $4-$component Dirac spinor ${\bf \psi} (1)$ by a $4 \times 4
\, = \, 16$ component spinor ${\bf \psi} (1,2)$ and required that it be
an eigenfunction of an operator which is simply the sum of the $h_D ^{ext}
(i)$ , and $U_C (1,2)$ .  I like to call this operator the
(two-electron) Dirac-Coulomb Hamiltonian and denote it by $H_{DC} (1,2)$
:
\begin{equation}
H_{DC} (1,2) \, = \, h_D ^{ext} (1) \, + \, h_D ^{ext} (2) \, + \, U_C
(r_{12}) .
\label{4.5}
\end{equation}
In this notation, Breit's requirement takes the form
\begin{equation}
H_{DC} {\bf \psi} (1,2) \, = \, E {\bf \psi} (1,2).
\label{4.6}
\end{equation}
The normalizable solutions of this equation were presumed to give a good
description of the discrete states of the helium atom.

\subsubsection{Transverse photon exchange and the Breit operator.}
Next, Breit took into account effects arising from the interaction of
the electrons with the quantized radiation field $ {\bf A}_T ({\bf x})$
by including the interaction which, in this context is given by
\begin{equation}
H_T \, = \, \Sigma e {\bf \alpha}_i \cdot {\bf A}_T ({\bf r}_i) \;
\label{4.7}
\end{equation}
Using second-order perturbation theory and dropping terms which
correspond to self-interaction, Breit argued that to a good
approximation the level shift $\Delta E$ arising from this
interaction would be given by
\begin{equation}
\Delta E \, = \, <\psi \mid U_B \mid {\bf \psi}>/<{\bf \psi} \mid {\bf \psi}>
\label{4.8}
\end{equation}
where $U_B$ is an operator which came to be known as the Breit operator,
\begin{equation}
U_B^{(1,2)} \, = \, -(e^2/4 \pi)({\bf \alpha}_1 {\bf \cdot} {\bf \alpha}_2 \,
 + \, {\bf \alpha}_1 {\bf \cdot \hat{r} \alpha}_2 \cdot {\bf \hat{r}})/2r,
 \quad ({\bf r \, = \, r}_1 - {\bf r}_2)
 \label{4.9}
 \end{equation}
 and ${\psi}$ is a normalizable solution of $(\ref{4.6})$.

 \subsubsection{Reduction to large components.}
 Finally, Breit used a procedure, which came to be known as ``a reduction
 to large components", to obtain from (\ref{4.6}) and ({\ref{4.8})
 the previously known forms of the operators which gave $\alpha^2 Ry$
 level shifts.  Recall first that in the standard representation of the
 Dirac matrices it is convenient to organize the four components of the
 Dirac wave function ${\psi}(1)$ of an electron into two Pauli
 spinors, ${\psi}_U (1)$, the first and second (or upper) components of
 ${\psi}$, and ${\psi}_L (1)$, the third and fourth (or lower) components of
 $\psi(1)$.  For
 a hydrogenic bound state ${\psi}_L$ is of order $(v/c) {\psi}_U$, so in
 this context the upper are the ``large" and the lower
 are the ``small" components of ${\bf \psi}(1)$, respectively.  One can
 similarly decompose a 16-component spinor ${\psi} (1,2)$ into
 upper-upper components ${\bf \psi}_{UU} (1,2)$, lower-upper components
 ${\bf \psi}_{LU} (1,2)$, etc.  Breit wrote down the four coupled
 equations for the ${\psi}_{XY}$ which follow from $(\ref{4.6})$ and
 from three of these obtained {\it approximations} for ${\psi}_{LU}$,
 ${\bf \psi}_{UL}$, and ${\psi}_{LL}$ in terms of ${\psi}_{UU}$,
 the large-large components.  When substituted in the remaining equation
 the result is an equation for ${\bf \psi}_{UU}$ which reduces to
 (\ref{3.3}) in the extreme n.r. limit and contains further terms
 which can be identified as the previously known spin-self-orbit, kinetic
 energy, and spin-independent contact interactions.  Further, on using
 the leading approximations to the small components (e.g. ${\psi}_{LU} (1,2))
  \, \approx \, ({\bf \alpha}_1 {\bf \cdot
 p}_1^{op}/2m) {\psi}_{UU} (1,2))$ in the expression (\ref{4.8}),
 Breit found that $\Delta E$ reduced to the expectation value with the
 nonrelativistic wave function of the orbit-orbit, spin-other-orbit, and
 spin-spin interactions mentioned above.

\subsection{\it Critique of the early work}

The most serious
flaw of Breit's work is the circumstance that the
zero-order equation, the Dirac-Coulomb equation (\ref{4.6}), has no
bound states; that is, ${H}_{DC}$ has no normalizable eigenfunctions!
This startling fact was first pointed out by Brown and Ravenhall,
some 20 years after Breit's work.\cite{brownandravenhall}  Its
validity is most easily seen by first turning off $U_C(r)$ in
${H}_{DC}$.  The resulting Hamiltonian does have normalizable
eigenfunctions, of the form ${\bf \psi}_{mn}^{(o)} \, = \,
u_m(1)u_n(2)$, with $u_m(1)$ and $u_n(2)$ denoting normalizable
eigenfunction of ${h}_D^{ext}(1)$ and ${h}_D^{ext}(2)$, with
enegies $\epsilon_m$ and $\epsilon_n$, respectively.  But
such product states are degenerate in energy with similar product
functions in which, say, "1" is in a continuum state with positive
energy $\epsilon_1 \, > \, m$ and "2" in a continuum state with
negative energy $\epsilon_2 \, < \, -m$ and $\epsilon_1+\epsilon_2 \, =
\, \epsilon_m+\epsilon_n$.  Since the sum $\epsilon_1+\epsilon_2$ can be
kept constant with $\epsilon_1$ increasing and $\epsilon_2$ decreasing,
the degeneracy is continuous.  Thus, turning on any interaction
which has non-zero matrix elements between positive- and negative-energy
eigenfunctions of $H_{CD}^{(o)}$, such as $e^2/r_{12}$, will lead to
a mixture of ${\bf \psi}_{mn}^{(o)}$ with a sea of non-normalizable states,
i.e. to ``continuum dissolution."\cite{Suchera}

The CD problem undermines {\bf both} the derivation of the correction
terms {\it not} associated with transverse photon exchange and the
derivation of (\ref{4.8}), which purports to take these into account.
However, suppose one overlooks the CD problem and i) considers the wave
function ${\bf \psi}^{(o)}(1,2)$ with components
\begin{equation}
{{\bf \psi}^{(0)}}_{LU}(1,2) \approx ({\bf \sigma}_1 \cdot {\bf
p_1}/2m){\bf \psi}^{(0)}_{UU}(1,2),
\label{4.10}
\end{equation}
etc., with ${\bf \psi}^{(0)}_{UU}(1,2)$ identified with the n.r. SP wave
function, as an approximation to an eigenfunction of an as yet
unspecified relativistic zero-order Hamiltonian, (ii) accepts the
formula (\ref{4.8}) and evaluates it with the function ${\bf
\psi}^{(0)}(1,2)$, as was done by Breit.  On carrying out some
integration by parts and being careful to keep nonvanishing surface
terms, one finds that, apart from the terms found by him, there
is an additional term, a spin-spin contact interaction between
electrons\cite{SucherFoley,SesslerFoley}:
\begin{equation}
U_{s-s}^{cont} = -(8 \pi /3)(e^2/4 \pi){\bf \sigma}_1 \cdot {\bf
\sigma}_2 \delta ({\bf r}) m^{-2}.
\label{4.11}
\end{equation}
This term is simply the counterpart of the Fermi contact interaction
between an electron and a proton, which had been known for many years
previously in the context of the hydrogen hyperfine structure.  It is
somewhat scandalous that it took so many years to include this effect in
fine-structure calculations in atomic physics.

In summary, Breit's theory of the helium atom was flawed both on
conceptual and technical grounds.  To a large extent the reason that
this was not noticed for more than two decades is that he got the
``expected answer" for terms previously known and that experiments were
not precise enough to measure level shifts of order $\alpha^2 Ry$ with
great accuracy.

A sign that something was badly amiss with the general theory should
have come from the following fact.  Breit briefly considered the
possibility of treating $U_B$ nonperturbatively by adding it to $U_C$ and
using as a starting point what I call the Dirac-Coulomb-Breit
(DCB) Hamiltonian:
\begin{equation}
H_{DCB}(1,2) = h_D^{ext}(1) + h_D^{ext}(2) + U_C(r_{12}) +U_B(r_{12})
\label{4.12}
\end{equation}
However, he soon noted that use of $U_B$ in second-order perturbation
theory led to strange results: The contribution from negative-energy
states is readily seen be of order $<U_B^2/4m>$ and since $U_B^2 \sim
U_C^2$ this is of order $\alpha^2 Ry$, as large as the first-order
contribution of $U_B$!  So the Breit operator always came with the caveat:
It was to be used only in first-order perturbation theory.

Nevertheless, it should be emphasized that Breit's work was a pioneering
example of the effort to derive effective potentials from quantum field
theory, apparently accomplished when the first paper on the general
quantum theory of fields by Heisenberg and Pauli was available only in
preprint form.

\section{Quantum field theory of two-electron atoms:  Foundations
and level-shifts to order ${\bf \alpha^3 Ry}$}

\subsection{\it Foundations}

In the 50's, motivated by the increasingly accurate experimental
results which were becoming available, e.g. for the splitting of the $2^3 P$
states and for the ionization energy, the foundation of
the theory of the level-structure of $He$ and $He$-like
ions was reexamined, independently by H. Araki\cite{Araki} and
me.\cite{JSucher} The starting point for both these studies was a
generalization of the two-body Bethe-Salpeter equation\cite{Salpeter}
to include an external field, an integral equation for a 16-component
spinor function ${\bf \psi}(x_1,x_2)$, of the form
\begin{equation}
{\bf \psi}(x_1,x_2) = -i \int K_a(x_1,x_3)K_b(x_2,x_4)G(x_3,x_4;x_5,x_6)
{\bf \psi}(x_5,x_6)d \tau .
\label{5.1}
\end{equation}
Here $d\tau = d^4x_3d^4x_4d^4x_5d^4x_6$ is the product of four-dimensional
volume elements, $K_a(x_1,x_3)$ is the external-field Feynman propagator for
electron "a", and $G$ is an interaction kernel, constructed from
two-particle-irreducible {\it x}-space Feynman graphs,
involving the emission and
absorption of one or more photons.  In these graphs, the electron propagators
are the $K_a$ and the photon propagators depend on the choice of gauge.
If one works in Coulomb gauge one can, by an extension of
a method of Salpeter,\cite{Salpeter} derive an exact equation for the spatial
part $\phi({\bf x}_1,{\bf x}_2)$ of the equal-times wave function
$\psi({\bf{x}}_1,t; {\bf{x}}_2,t)$ associated with a stationary state of
energy $E$,
\begin{equation}
{\bf \psi}({\bf x}_1,t; {\bf x}_2,t) = \phi({\bf x}_1,{\bf x}_2) e^{-iEt} .
\label{5.2}
\end{equation}
This equation, which I shall not write down, is still linear but involves
the eigenvalue {\it E} nonlinearly.  On neglecting effects
associated with electron-positron pairs and transverse photons in this
equation, one arrives at what I call the ``no-pair Coulomb-ladder (NPCL)
equation":
\begin{equation}
H_{++} \phi_C ({\bf x}_1,{\bf x}_2) = E \phi_C({\bf x}_1,{\bf x}_2).
\label{5.3}
\end{equation}
Here
\begin{equation}
H_{++} = h_D^{ext}(1) + h_D^{ext}(2) + L_{++}U_C(r_{12})L_{++}
\label{5.4}
\end{equation}
is the NPCL Hamiltonian and
\begin{equation}
L_{++} \equiv L_+(1)L_+(2)
\label{5.5}
\end{equation}
is the product of external-field positive-energy projection operators
$L_+(i)$, defined by
\begin{equation}
L_+(i) = \Sigma \mid \phi_n(i)><\phi_n(i) \mid.
\label{5.6}
\end{equation}
Here the sum is over the positive-energy eigenstates $\phi_n$
(bound and continuum) of $h_D^{(ext)}(i)$.  The NPCL wave functions
$\phi_C$ of physical interest satisfy the constraint
\begin{equation}
L_+(i)\phi = \phi  \quad     (i = 1,2).
\label{5.7}
\end{equation}
This is consistent with (\ref{5.3}) since the $L_+(i)$ commute with
$H_{++}$,
\begin{equation}
[L_+(i), H_{++}] = 0.
\label{5.8}
\end{equation}
The nomenclature "NPCL" arises from the fact that (\ref{5.3}) effectively
sums all time-ordered diagrams which involve repeated Coulomb scattering
of the electrons in the presence of the external field, with no
electron-positron pairs present in intermediate states.

The NPCL equation (\ref{5.3}) has the following features:
(a) Unlike the DC equation (\ref{4.6})
it does not suffer from continuum dissolution, because when
$U_C(r_{12})$ is turned on the projection operator $L_{++}$ prevents
the mixing referred to above. (b) It is exact to all orders in the
external field strength. (c) Because of the condition (\ref{5.7})
it can be reduced without approximation to $SP$ form, with the eigenvalue
{\it E} still appearing linearly. (d) This relativistic $SP$
equation reduces to (\ref{3.2}) in the extreme $NR$ limit and contains all the
$\alpha^2 Ry$ corrections not associated with transverse photons.

The NPCL wavefunction can therefore be used as the basis of a
perturbative treatment of the four-dimensional equation (\ref{5.1}),
as developed in Ref. 11, to compute effects associated with
transverse-photon exchange and virtual pairs.  In particular, the
leading effects of transverse-photon exchange, of order $\alpha^2 Ry$,
can be incorporated by simply adding $U_B$ to $U_C$ in (\ref{5.4}):
\begin{equation}
H_{++}' = h_D^{ext}(1)+h_D^{ext}(2) + L_{++}[U_C(r_{12})+
U_B(r_{12})]L_{++}).
\label{5.9}
\end{equation}
The corresponding equation
\begin{equation}
H_{++}' \phi_C' = E' \phi_C'
\label{5.10}
\end{equation}
has, like (\ref{5.3}), a clear-cut origin in field theory: It is obtained
from (\ref{5.1}) by keeping the one-photon exchange graph in the
interaction kernel \underline{G}, but neglecting the time-component
$k_0$ in the factor $(k^2 + i \epsilon)^{-1}$ appearing in the
transverse-photon propagator.  It can be regarded as a QFT-based
replacement for the patchwork approach to fine structure.
The eigenvalues $E'$ incorporate {\bf all} fine-structure effects of
order $\alpha^2 Ry$ and of course some of the effects of order
$\alpha^3 Ry$ and higher.

The basic difference between the modern approach and the earlier one
is not the four-dimensional formalism, which is mainly a convenient
tool and can to a large extent be dispensed with, but that
QED incorporates the ideas of Dirac's hole theory rather than the
Dirac's one-electron theory.  The appearance of the projection
operators is the mathematical expression, in the present context,
of the idea that the negative-energy states are filled and transitions
to them are forbidden by the Pauli exclusion principle.

\subsection{$\alpha^2 Ry$ and $\alpha^3 Ry$ level-shifts}

\indent Analysis of (\ref{5.4}) shows that the low-lying
eigenvalues have the form
\begin{equation}
E = 2m + W + \delta W_C
\label{5.11}
\end{equation}
where W is the n.r. eigenvalue of (\ref{3.2}) and
\begin{equation}
\delta W_C  = \delta W_C^{(2)} + \delta W_C^{(3)} + ...
\label{5.12}
\end{equation}
The first term in (\ref{5.12}), of order $\alpha^2 Ry$, is defined by
\begin{equation}
\nonumber \delta W_C^{(2)} = <\phi \mid {-{``\bf p}_1^4/8m^3" +
\nonumber (\pi \alpha /2m^2) [Z\delta ({\bf r}_1) - \delta({\bf r}_{12})]} +
\nonumber{1 \leftrightarrow 2} \mid \phi>
\end{equation}
\begin{equation}
+ <\phi \mid{(\alpha/4m^2){\bf \sigma}_1
\cdot (Z {\bf r}_1 \times {\bf p}_1
/r_1^3 - {\bf r}_{12} \times {\bf p}_1/r_{12}^3)} +
{1 \leftrightarrow 2} \mid \phi>
\label{5.13}
\end{equation}
and is the sum, for each electron, of a kinetic energy correction,
a spin-independent contact interaction, and a spin-self-orbit interaction.
The latter terms are natural extensions of those found from the Dirac
equation in the one-electron case, if one recognizes that electron
$``2"$ serves as the source of an electric field just as well as the nucleus.
The quotes on the ${\bf p}^4$ term serve as a reminder that this term
must be evaluated as
\begin{equation}
-<{\bf p}_1^2 \phi \mid {\bf p}_1^2 \phi>/8m^3,
\label{36}
\end{equation}
to avoid double counting.\cite{SucherFoley}  The second term in (\ref{5.12}),
defined by
\begin{equation}
\delta W_C^{(3)} = -(\alpha^2/m^2)(\pi/2 + 5/3) <\phi \mid \delta ({\bf
r}_{12}) \mid \phi>,
\label{5.13a}
\end{equation}
is an s.i. contact term of order $\alpha^3 Ry$, arising directly from
(\ref{5.4}).

The remaining corrections of order $\alpha^2 Ry$ and another correction of
order $\alpha^3 Ry$ may be obtained by a similar analysis starting from
(\ref{5.9}) or, equivalently, by evaluating
\begin{equation}
\delta E_B = <\phi_C \mid U_B \mid \phi_C>
\label{5.14}
\end{equation}
with a sufficiently accurate approximation to $\phi_C$.  The result is
\begin{equation}
\delta E_B = \delta W_B^{(2)} + \delta W_B^{(3)} + ...
\label{5.15}
\end{equation}

\noindent The first term in (\ref{5.15}) which comes from reduction of the
Breit operator, using approximations such as (\ref{4.10}), is given by
\begin{equation}
\delta W_B^{(2)} = < \phi \mid U_{o-o} + U_{s-o-o} + U_{s-s} \mid \phi>
\label{5.16}
\end{equation}
with
\begin{equation}
U_{o-o} = (\alpha/2m^2)r_{12}^{-1}[{\bf p}_1{\bf \cdot p}_2+{\bf \hat{r}
\cdot} ({\bf \hat{r} \cdot p}_1) {\bf \cdot p}_2]
\label{5.17}
\end{equation}
a form of the orbit-orbit interaction, 
\begin{equation}
U_{s-o-o} = -(\alpha/2m^2)r_{12}^{-3}[{\bf \sigma}_1 {\bf \cdot} {\bf r}_{12}
\times {\bf p}_2 + 1 \leftrightarrow 2]
\label{5.18}
\end{equation}
the spin-other-orbit interaction, and $U_{s-s}$ the full spin-spin
interaction, including the contact term (\ref{4.11}):
\begin{equation}
U_{s-s} = (\alpha/4m^2)[r_{12}^{-3}({\bf \sigma}_1 {\bf \cdot \sigma}_2-
3{\bf \sigma}_1 {\bf \cdot {\hat r} \sigma}_2 {\bf \cdot {\hat r}}) - (8 \pi/3)
{\bf \sigma}_1 {\bf \cdot \sigma}_2 \delta({\bf r})].
\label{5.19}
\end{equation}
The second term in (\ref{5.15}) is a spin-dependent partner of (\ref
{5.13a}) and contributes to fine-structure splitting in order $\alpha^3 Ry$:
\begin{equation}
\delta W_B^{(3)} = -(\alpha^2/m^2)(4/3)(\pi/2+1)<\phi \mid \mid
{\bf \sigma}_1 \cdotå {\bf \sigma}_2 \delta({\bf r}_{12}) \mid \phi>.
\label{5.20}
\end{equation}
The corrections (\ref{5.13a}) and (\ref{5.20}) are both associated with the
fact that the usual approximations to the LU, UL and LL components of the
16-component wave function are not sufficiently accurate at short distances,
for the purpose at hand.
                               
The computation of the remaining terms of order $\alpha^3 Ry$ requires a
return to (\ref{5.1}) and i) keeping $k_0$ in the transverse photon propagator
(this corresponds to taking into account the Coulomb interaction during the
exchange of one transverse photon), ii) keeping the irreducible two-photon
exchange graph in the kernel $G$, and iii) keeping graphs associated with
radiative corrections, the latter of course include self-energy effects,
associated in hydrogen with the Lamb-shift.  In $He$, such effects were
first estimated by Kabir and Salpeter.\cite{KabirSalpeter}  So much for
the 50's.  Later work in this area will be mentioned in Sec. 8.
	
\section{Radiative decays of helium: Enter the Sun once more}
Because spin-dependent forces are weak if Z is not too large, the
states of $He$ or low-Z $He$-like ions may be classified into singlet and
triplet states, corresponding to total spin 0 and 1, respectively.
The lowest lying triplet S-state, described as $2^3S_1$, lies above
the $1^1S_0$ ground state and is expected to be very long-lived,
because of the form of the (nonrelativistic) radiation operator for
one-photon decay, viz.
\begin{equation}
R_{rad} = (e/ \sqrt{2 \omega}) \Sigma({\bf p}_i/m -i{\bf s}_i \times {\bf k}/m)
\newline \cdot {\bf \epsilon}^*exp(-i{\bf k \cdot r}_i).
\label{5.21}
\end{equation}
The orbital part of $R_{rad}$ vanishes between $L = 0$ states and in the
dipole approximation the spin part of $R_{rad}$ becomes proportional to the
total spin {\bf S}.  So the matrix element for the would-be magnetic
dipole transition ($\Delta J = \pm 1$, 0; no parity change):
\begin{equation}
2^3S_1 \to 1^1S_0 + \gamma
\label{5.22}
\end{equation}
is doubly vanishing in this approximation:  The matrix element of {\bf
S} is zero between a triplet and a singlet spin wave function and the
spatial parts of the initial and final states, belonging to different
eigenvalues of (3), are orthogonal. 

What then is the rate for this decay?  This problem was apparently first
studied in 1940 by Breit and Teller (BT) who, motivated by questions in
astrophysics, wrote a pioneering paper on the $1 \gamma$ and $2 \gamma$ decays
of $H$ and $He$.\cite{Breit}  They estimated that
$R(2^3S_1 \to ^1S_0 +\tau) \approx 10^{-24} R$ where $R$
denotes the rate expected for an E1 decay of unit
oscillator strength $(\approx 10^9 s^{-1})$, corresponding to a lifetime
for the $1 \gamma$ decay of $10^8$ years!  This was the last word on the
subject for almost three decades, when the Sun enters our story once more.
In 1969 it was noted by A. H. Gabriel and C.
Jordan\cite{GabrielJordan} that certain lines seen in the soft x-ray
spectrum of the solar corona correspond in wavelength to the transition
(\ref{5.22}) in certain $He$-like ions, ranging from C V to Si XIII,
and suggested that these transitions were in fact responsible for the
observed lines.  However, they noted that the intensities for these lines
would be extremely small, if the result of BT were simply scaled up,
even with a high power of $Z$.

Soon thereafter H. Griem pointed out that BT vastly underestimated the
rate for (\ref{5.22}) in $He$ and made an estimate, patterned on the (largely)
correct calculation of BT for the hydrogenic decay $2^{1/2}S_{1/2} \to
2^{1/2}S_{1/2} + 1 \gamma$, which indicated that the Gabriel-Jordan
interpretation was plausible.\cite{Griema}  A similar estimate was made by
G. Feinberg.\cite{Feinberga}  At about the same
time the whole subject was brought
down to Earth, so to speak, with the development of beam foil spectroscopy
and the terrestrial measurement of such rates, beginning with that of
(\ref{5.22}) in $Ar^{16+}$ by Marrus and Schmieder.\cite{MarrusSchmieder}
This raised interest in the whole subject and led to the development of a
general QED based theory of such decays.\cite{FeinbergSucher,Drakea}

A correct starting point for the calculation of one-photon decays in
$H$-like ions which includes relativistic effects is the equation
\begin{equation}
M_1 = <{\bf \psi}_f \mid {\bf \alpha \cdot \epsilon ^*}e^{-i{\bf k \cdot
r}} \mid {\bf \psi}_i>.
\label{5.23}
\end{equation}
where the ${\bf \psi}$ 's are normalized solutions of (\ref{4.2}).  This is
also what one would write down in Dirac's one-electron theory and it
yields the correct leading terms for all such decays.  Thus the
counterpart of (\ref{5.23}) for a two-electron system is
\begin{equation}
M_2 = \Sigma <{\bf \psi}_f(1,2) \mid {\bf \alpha}_i {\bf \cdot \epsilon
^*} e^{-i{\bf k \cdot} {\bf r}_i} \mid {\bf \psi}_i(1,2)>,
\label{5.24}
\end{equation}
where now the ${\bf \psi}$'s are solutions of (\ref{5.3}),
which is according to the above discussion a QED-based analogue of
(\ref{5.23}).  However, it turns out that (\ref{5.24}) does
\underline{not} give the full leading terms for the M1 decays in question:
One must go back to QED and include time-ordered graphs in which, e.g.,
the Coulomb field of electron ``1" or the nucleus creates a virtual
$e^- - e^+$ pair
and the positron annihilates with electron ``2" to create the emitted photon.
{\it So field theory rears its head again}.  The final result is
\begin{equation}
M(2^3S_1 \to 1^1S_0 + \gamma) \approx 2i \Sigma_{fi} T_{fi}
\end{equation}
where
\begin{equation}
T_{fi} = <\phi_f \mid {\bf p}_1^2/3m^2 + k^2 r_1^2/12-r_1U'(r_1)/6m \mid
\phi_i>
\end{equation}
and
\begin{equation}
\Sigma_{fi} = <{\bf \chi}_f \mid \sigma_1 {\bf \cdot k} \times {\bf \epsilon}^*/m
\mid {\bf \chi}_i>
\end{equation}
with ${\bf \chi}_i$ and ${\bf \chi}_f$ the triplet and singlet spin wave
functions.  The corresponding decay rate is then found to be
\begin{equation}
R_{He-like} = (2/3)(8 \alpha k^3/m^2) \mid T_{fi} \mid^2
\end{equation}
The evaluation of $T_{fi}$ for large $Z$ was first carried out by Drake using
many-parameter variational wave functions and an expansion in powers of
$1/Z$.  Because of early discrepancies in the cases of $Ar^{16+}$ and
$Cl^{15+}$, further work to estimate and include radiative, recoil,
retardation and relativistic corrections were carried out by various
workers.\cite{Sucher}  It turns out that for $Z$ not
too small, larger than say 15,
these results can be put together to write a cooperative explicit
formula:\cite{Sucher}
\begin{equation}
R_{He-like} = (2/3) \alpha (\alpha Z)^{10} (m/972) [1-4.10/Z + 6.7/Z^2 +
1.07(\alpha Z)^2]
\end{equation}
which could, conservatively, be expected to be
accurate to better than $2 \%$ in the
range $15 < Z < 40$.  Comparison with data ranging from sulfur (Z = 16) to
krypton (Z = 36), for which the lifetime ranges from 700 ns to .2 ns, a
factor of more than 3000, indeed showed agreement of theory and experiment
to within a few percent in all cases.  When one includes $He$, for which the
lifetime of the $2^3S_1$ state is about $10^4s$,\cite{Woodward} albeit
with a rather
large ($30 \%$) error, and evaluates $T_{fi}$ with a Hylleraas type
of wave function, the agreement extends over a range covering 14 orders of
magnitude.  This was surely a triumph of QED in a rather unusual domain.
Again, recent work in this area will be noted in Sec. 8.

\section{Retarded van der Waals forces and Rydberg states of helium.}

Van der Waals forces are, in the first instance, the (spin-independent)
electromagnetic forces which act between atoms at separations
{\it R} large compared to atomic size {\it a}.  Two years
after the development of quantum mechanics, S.C. Wang and later F.
London showed that the electrostatic interaction between the
constituents of two neutral atoms gives rise to an effective interatomic
potential of the form $U = -C/R^6$ for $R \gg a$
(often called the London potential).\cite{Wang}
But it took two decades after the
development of QED, for H.B.G. Casimir and D. Polder to point out that
the inclusion of the effects of transverse photon exchange between the
electrons leads to a potential which falls of as $R^{-7}$ at very large
separations, of the order of the maximum wavelength for electric dipole
excitation.\cite{Casimir}  Inclusion of transverse
photon exchange corresponds, in the
framework of QFT, to taking into account the retarded character of
electromagnetic interactions in CED.  Thus one refers to the potential
at very large $R$ either as the Casimir-Polder potential or as the
retarded van der Waals potential.  Unfortunately, because the potential
is very small in this region, direct experimental detection of the
retarded character of atom-atom van der Waals forces has proved to be
elusive.

On general grounds, one expects that related retardation effects
manifest themselves between a polarizable neutral system $A$ and a
charged system $B$, or between two charged systems $A$ and $B$.  However
in these cases, such effects will only be corrections to the
dominant forces, corresponding to the electric dipole potential $U_{dip}
= - {\bf \alpha}_A^{el} e_B^2/2R^4$ and the Coulomb potential
$U_C$, respectively.  It was pointed out by L. Spruch that precise
measurement of the fine structure of Rydberg levels in helium might be
able to detect such corrections and S. Lundgren and co-workers undertook
a series of experiments to detect such effects, specifically on the $n =
10$ levels.\cite{Drake2}  This stimulated a great deal
of further work on the QED aspects by theorists including Spruch and
J.F. Babb, and C.K. Au, G. Feinberg and myself.\cite{Drake2}
There is only one point I will elaborate on in this regard, by way
of making a connection with the beginning of the modern theory of
helium.  G.W.F. Drake, by using a doubled set of Hylleraas-type wave
functions and modern computer power, has extended the variational
methods pioneered by Hylleraas so long ago to achieve unprecedented
accuracy in computing the eigenvalues and eigenfunctions of the
nonrelativistic Hamiltonian (3).\cite{Drake2}
R. Drachman has achieved very high
accuracy for the large {\it n}, large {\it l} states, by using a
systematic perturbation expansion in the residual interaction
$e^2(1/r_{12} -1/r_1)$ of the outer electron with the
core.\cite{Drake2}  Thus helium now
approaches hydrogen in having an exactly-solvable zero-order problem.
As a result, the calculation of relatively high-order relativistic and QED
type corrections in helium and helium-like ions has become much more
meaningful than in earlier decades, as emphasized below.

\section{Some recent developments; Concluding remarks}

As described in Sec.~5, the order ${\bf \alpha}^3 Ry$ corrections to the
low-lying ${\bf \alpha}^2 Ry$ fine structure of helium were
calculated long ago.  Increased experimental precision and the
possibility of using $He$ fine-structure for obtaining values
of the fine-structure constant competitive with other methods, has led
to improvements on the earlier calculations and to the computation of
still higher-order effects.  Helium has also been used as a testing
ground for techniques vital for studying atoms with more than two
electrons, such as relativistic many-body perturbation theory (MBPT),
configuration interaction (CI) methods, and multi-configuration
Dirac-Fock (MCDF) methods.

A major step in the direction of going to higher orders was taken by
M.H. Douglas and N. Kroll, who, using the methods of Ref. 11, obtained
formulas for fine-structure splitting of order $\alpha^4 Ry$.\cite{DouglasKroll}
Recently this work has been improved and extended by
several theoretical groups.  T. Zhang and G.W.F. Drake\cite{ZhangDrake}
and Eides {\it et al.}\cite{Eides} have verified the correctness of the somewhat
phenomenological treatment of radiative corrections by Douglas and
Kroll.  Z.-C. Yan and Drake \cite{YanDrake} have computed the
fine-structure of the $2^3P$ state in {\it He} and low-$Z$ {\it
He}-like ions to a few parts in $10^9$, obtaining good agreement with
recent high precision measurements\cite{Shiner} for $Z = 3$,5, and 9
but not for $Z = 2$ and 4.  Complementary calculations for higher $Z$ have
been carried out by M.H. Chen {\it et al.},\cite{Chen} using
relativistic CI methods, and by D.R. Plante {\it et al.}\cite{Deplante},
using relativistic MBPT methods.  In a heroic computational
effort, T. Zhang has obtained corrections of order ${\alpha}^4ln{\alpha}Ry$
to energy levels and corrections of order ${\alpha}^5 ln{\alpha}Ry$ and
${\alpha}^5 Ry$ to fine-structure splitting.\cite{Zhang} These will be
important in using helium fine structure to obtain an independent
determination of ${\alpha}$; a recent measurement by Myers {\it et
al.}\cite{Myers} of the fine-structure splitting of the $2^3P$ splitting
for Z = 7 is of sufficient accuracy to be sensitive to such corrections.

Much progress has been made in recent years in developing new techniques
for the accurate calculation of self-energy corrections in $He$ and
$He-$like ions, i.e. corrections which are the counterpart of
what is referred to as the Lamb shift in $H$and $H$-like ions. A recent
example, where further references may be found, is the work of Persson
{\it et al.}\cite{Persson}, in which good agreement is found with
experimental results of Marrs {\it et al.}\cite {Marrs} on several
$He$-like ions, ranging from Z = 32 to Z = 83.

With regard to decay rates, W.R. Johnson {\it et al.}\cite{Johnson} have carried
 out extensive relativistic calculations of radiative decay rates in the
helium isoelectronic sequence, using the no-pair Hamiltonian (26) as a
starting point.  These calculations are so accurate that
disagreement with experimental values could be interpreted as associated
with radiative corrections to decay rates, a relatively unexplored
subject.  Very recently, P. Indelicato\cite {Indelicato} has extended the
MCDF method to the calculation of the forbidden magnetic-dipole decays,
discussed in Sec. 6.

As a final example, in a quite different direction, of the way
in which helium engages the attention of physicists, I refer you to a
paper by Deilamian {\it et al.}\cite{Deilamian}, who have made a search
for $He$ states which are forbidden by the Pauli antisymmetry principle, by
way of putting an upper bound on its hypothetical violation.

\indent The story goes that when a mathematics professor was asked by a
student ``What is the use of number theory?", he
replied: ``Number theory is useful because you can get a Ph.D. with it."
As one may infer from the above saga, helium has
certainly been useful for getting a doctoral degree, and its uses in the
laboratory and in providing impressive views of football stadiums are
well known.  In a more serious vein, helium, together with its ionic partners,
is likely to continue
for a long time to come its unique role as a laboratory for the application
of quantum field theory and the techniques of mathematical physics to a
system of fundamental physical interest. It provides a unique
setting for the study of the electron-electron interaction, the electron
being one of the few observed particles which are still usefully regarded as
elementary.  While helium may be inert chemically, it remains highly active
intellectually.

\begin{center} {\bf Acknowledgments} \end{center}

This work was supported in part by the U.S. National Science Foundation.
I am grateful to I. Feranchuk, A. Gazizov, and O. Shadyro for their
hospitality and assistance during my stay in Minsk.

\end{document}